# Combining segregation and integration:
# Schelling model dynamics for heterogeneous population

Erez Hatna[1], Itzhak Benenson[2]

[1]Center for Advanced Modeling (CAM), Johns Hopkins University, USA

[2]Department of Geography and Human Environment, Tel Aviv University, Israel

**Abstract**

The Schelling model is a simple agent-based model that demonstrates how individuals' relocation decisions can generate residential segregation in cities. Agents belong to one of two groups and occupy cells of rectangular space. Agents react to the fraction of agents of their own group within the neighborhood around their cell. Agents stay put when this fraction is above a given tolerance threshold but seek a new location if the fraction is below the threshold. The model is well-known for its tipping point behavior: an initial random (integrated) pattern remains integrated when the tolerance threshold is below 1/3 but becomes segregated when the tolerance threshold is above 1/3.

In this paper, we demonstrate that the variety of the Schelling model's steady patterns is richer than the segregation–integration dichotomy and contains patterns that consist of segregated patches for each of the two groups, alongside patches where both groups are spatially integrated. We obtain such patterns by considering a general version of the model, in which the mechanisms of the agents' interactions remain the same, but the tolerance threshold varies between the agents of both groups.

We show that the model produces patterns of mixed integration and segregation when the tolerance threshold of most agents is either below the tipping point or above 2/3. In these cases, the mixed patterns are relatively insensitive to the model's parameters.

# 1. Introduction

The Schelling model of segregation was introduced by Thomas Schelling in the late 1960s (Schelling 1969, 1971, 1974, 1978). Schelling devised the model in order to demonstrate how individuals' relocation decisions entail global segregation. Schelling noted that his abstract model could reflect different spatial phenomena, but his main concern was the residential segregation of blacks and whites in United States cities (Schelling 1969, p488). In this interpretation, the model consists of households that make residential decisions based on the ethnic composition of neighborhoods.

Using today's terminology, the Schelling model is an agent-based model. Agents belong to one of two groups and occupy cells of rectangular space. A cell can either be empty or occupied by a single agent. Agents react to the fraction **f** of *friends* (i.e. agents of their own group) in the local neighborhood around their cell. An agent is satisfied with its cell when the fraction of friends is above the threshold **F**. However, an agent is dissatisfied when the fraction is below **F** (**f** < **F**), and consequently the agent tries to relocate to an empty cell where the fraction of friends within the cell's neighborhood is satisfactory (**f** ≥ **F**). **F** is the minimum fraction of friends required by an agent to be satisfied, and is known as the *tolerance threshold.*

A wide spectrum of formal representations of the Schelling mechanism exhibit a well-known tipping point behavior: An initial random pattern remains, in time, integrated for **F** < **F**$_{critical}$, while it converges to segregation for **F** ≥ **F**$_{critical}$. The tipping point **F**$_{critical}$ is about 1/3, indicating that a relatively weak individual tendency to segregate is sufficient for global segregation. Schelling used the moderate value of **F**$_{critical}$ to demonstrate that residential segregation in cities could form even when all of the individuals are willing to live within integrated neighborhoods and no ethnic discriminatory mechanism or economically-induced segregation exist (Schelling 1971).

Schelling's simplistic mechanism of residential relocation constitutes the core of more comprehensive models of residential dynamics (Ellis et al. 2012). Thus, it is important to reveal the spectrum of patterns that it can generate. In this paper, we demonstrate that the model's repertoire is not limited to merely integration (Figure 1a) or segregation (Figure 1b), but also include intermediate patterns of mixed integration and segregation. We refer to them as *mixed patterns.*

In the context of this study, we define mixed patterns as patterns that contain segregated patches for each population group as well as patches where both groups coexist (Figure 1c).

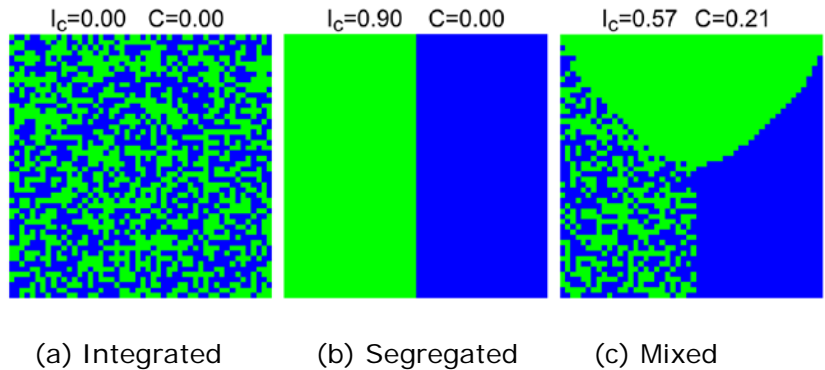

(a) Integrated  (b) Segregated  (c) Mixed

Figure 1: Patterns of blue and green agents. The indices are described below

We obtain mixed patterns by introducing variation in the agents' tolerance thresholds. The idea is to consider a population where part of the agents have tolerance thresholds below the tipping point (**F< F$_{critical}$**) while others have tolerance thresholds above (**F ≥ F$_{critical}$**). We study different distributions of tolerance thresholds and identify the ones that produce mixed patterns.

The paper has the following structure: Section 2 demonstrates the relevance of mixed patterns to the spatial ethnic patterns of cities. Section 3 presents a brief review of studies regarding the Schelling model. Section 4 includes a formal description of the model and Section 5 describes the tolerance distributions used in the investigation of the model. The indices that are used to characterize the patterns are described in Section 6, followed by the results in Section 7. We summarize our findings in Section 8.

## 2. Urban patterns of ethnicity

The ethnic segregation in US cities seems to have weakened since the 1970s (Glaeser and Vigdor 2012). However, the 2010 US census indicated that ethnic segregation is still prevalent in many cities (Logan and Stults 2011).

A visual assessment of the ethnic residential patterns of cities uncovers intricate patterns of segregation and integration, which are similar to the aforementioned mixed patterns. Figure 1 depicts two examples: Chicago and New York. The figure includes maps for the fractions of Black, Hispanic, White and Asian populations, alongside a map of the Shannon entropy index, which is calculated on the basis of these fractions[1].

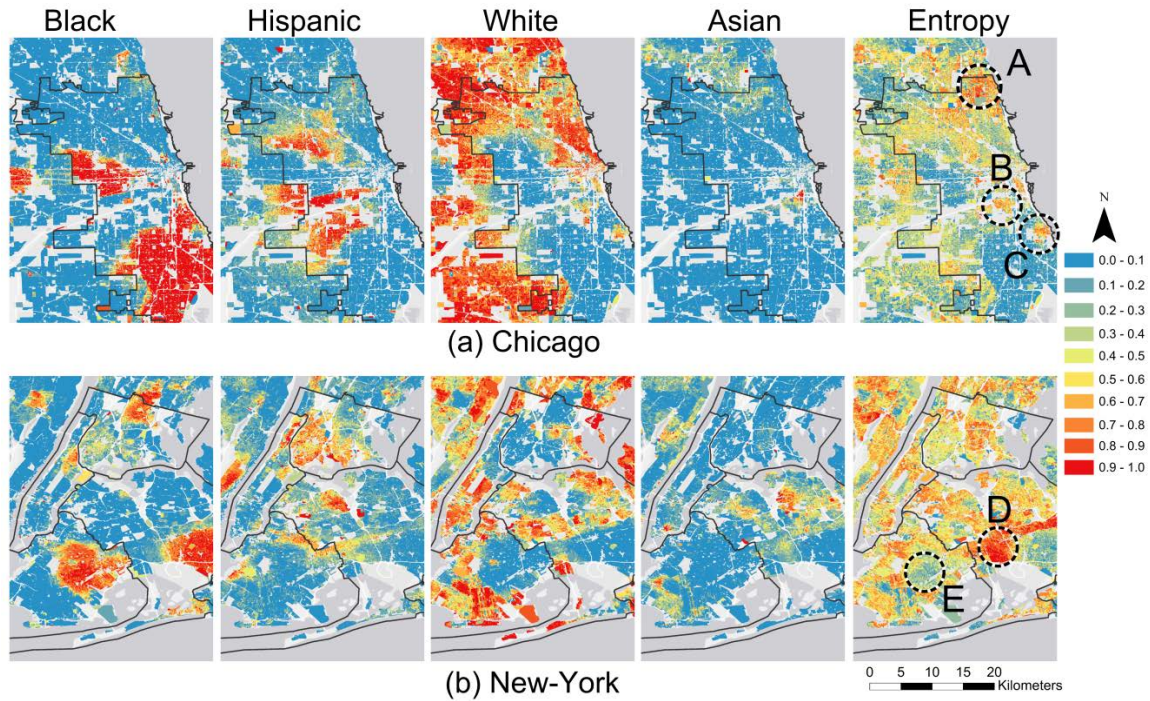

Figure 2: The fraction of Black, Hispanic, White and Asian adults within the census blocks in 2010, and the Shannon entropy index for Chicago and New York. The color legend is used for the population fractions as well as for the Shannon entropy index.

The city of Chicago contains large segregated areas of Black, Hispanic and White populations and a small segregated area of Asian population (Figure 2a). However, according to the entropy map, not all areas are segregated. The entropy map depicts areas that are populated by a variety of ethnic groups. To name a few, these include communities in the north, such as Rogers Park, West Ridge, Edgewater and Uptown (marked as A), as well as other communities, such as Bridgeport (marked as B) and Hyde Park (marked as C).

New York, is less segregated than Chicago (Figure 2b). Examples of New York neighborhoods that are ethnically diverse are Richmond Hill, Woodhaven and Ozone Park in Queens (marked as D), while Brooklyn neighborhoods, such as East Flatbush and Remsen Village (marked as E) are segregated. Ethnic patterns that are composed of mixed integration and segregation exist in many other US cities, as well as in Israeli cities (Hatna and Benenson 2012).

---

[1] The Shannon entropy index also includes the census "Other" category, which is not shown in Figure 2

## 3. The Schelling model

*3.1. Schelling's original studies*

Schelling introduced the original version of the model in 1969 (Schelling 1969). He referred to it as the *linear model* because it has the form of a one-dimensional array. The array is populated by "stars" ("+") and "zeros" ("0") representing agents belonging to two groups. An agent is dissatisfied when the fraction of friends in the four nearest cells of each side is below the tolerance threshold. If so, the agent relocates to the nearest position where the fraction of friends is above the threshold. Because all cells are occupied, the agent is inserted into its new position by shifting the position of the other agents. The model is updated in discrete steps by considering one agent at a time using a predefined order. Schelling illustrated that for **F**=0.5, an initial random pattern segregates into a long sequence of stars and zeroes.

In a later publication (Schelling 1971), Schelling presented the two-dimensional version of the model. In this model, agents are not shifted sideways in order to make room for relocating agents; instead, Schelling introduced empty cells as potential destinations for relocating agents. Agents move to the nearest empty cell that has a sufficient fraction of friends when dissatisfied with their local 3-by-3 neighborhood. In that paper, Schelling revealed for the first time that **F**$_{critical}$ is essentially below the intuitive value of 1/2 and is close to 1/3. That is, for **F** < 1/3, an initially random residential pattern remains random-like while for **F** ≥ 1/3, the pattern converges to a state of global segregation (Schelling 1971, p158).

*3.2. Later studies*

The body of literature on the Schelling model has been accumulating since the 1970s, and thus, it is far too large and fragmented to allow for a comprehensive review within this paper. Hence, we limit our review to relevant publications.

Benenson and Hatna (2011) were the first to reveal that the Schelling model can generate patterns where integration and segregation co-exist. More specifically, they demonstrated that when the two groups are of different sizes and **F** < **F**$_{critical}$, the model produces a segregated patch for the majority group while the rest of the area is occupied by both groups. In a later study, Hatna and Benenson (2012) showed that the model generates similar patterns when the tolerance thresholds are group-specific. Under these settings, the model generates similar patterns when the tolerance threshold of one group is above $F_{critical}$, but moderate while the tolerance threshold of the second group is below $F_{critical}$. It is worth noting that the patterns generated in these two studies are simpler than the mixed patterns that are the focus of this paper.

Xie and Zihaou (2012) were the first to study a Schelling model where each agent holds a personal preference concerning the ethnic composition of neighborhoods. Their model is an extension of the model developed by Bruch and Mare (2006, 2009) where all white and black agents have the same utility functions. They based the agents' tolerance regarding neighborhood composition on a US survey that revealed large heterogeneity in whites' tolerance for black neighbors. The model generates lower level of racial residential segregation when compared to the case of common racial tolerance. The authors mention that tolerant agents reside within less segregated neighborhoods, but did not indicate whether the model produces patterns of mixed integration and segregation.

Vinkovic and Kirman (2006) provide a general insight into the relation between rules and patterns by considering a continuous analog of the Schelling model. The authors describe two fundamental types of relocation rules. The first includes rules that enable relocation to a *better location only*. These rules generate patterns that stall with an essential fraction of discontent agents who are unable to find a better location. The second rules enable *relocation to cells of the same utility*. They generate patterns that do not stall, even when all agents are satisfied. In time, the patterns converge to a state where the statistical characteristics of the patterns (such as level of segregation) do not change.

Vinkovic and Kirman (2006) demonstrate that the patterns are very sensitive to parameters when the rules allow for relocation to better cells only. For example, the size of the clusters depends on the fraction of empty cells. In contrast, rules that allow for relocation to cells of the same utility produce patterns that are far less sensitive to the model settings. For example, all segregated patterns consist of two clusters (one for each group) when a torus grid is used.

In this study, we follow Vinkovic and Kirman's (2006) insight and employ relocation rules that allow for agent's movement between cells of the same utility. However, we restrict such movements to satisfied agents. Dissatisfied agents migrate to the better cell only.

## 4. Model rules

The rules of the model used in this study follow Hatna and Benenson (2012). Space is represented by a **N × N** array of cells on a *torus.* A cell is either vacant or populated by a single agent. We denote an agent as **a,** a cell as **c,** and the Moore **n × n** neighborhood of **c**, excluding **c** itself, as **U(c).** Each agent **a** belongs to one of two color groups: blue or green, and has a personal tolerance threshold $F_a$. An agent considers other agents of the same color as belonging to its own group; we refer to them as *friends*. Agent **a** can calculate the fraction of friends $f_a(c)$ among all agents located in **U(c)**. Hence, it ignores the vacant cells of **U(c)**.

Agent **a** evaluates the *utility* $u_a(c)$ of cell **c** within a non-empty neighborhood according to the fraction of friends and its personal tolerance threshold $F_a$:

$$u_a(c) = \min\{f_a(c), F_a\}/F_a \text{ if } F_a > 0 \quad (1)$$

$$u_a(c) = 1 \text{ if } F_a = 0$$

The utility of a cell in an empty neighborhood is defined as 0.

According to (1), the utility varies on [0,1]. The agent is satisfied with **c** if the fraction of friends within the **U(c)** is $F_a$ or higher. In this case, $u_a(c)=1$. If the fraction of friends in **U(c)** is lower than $F_a$, then the $u_a(c)=f_a(c)/F_a$; that is, the more friends the better.

Agent **a**, located in cell **h**, performs its relocation decision in two steps:

Step 1: The agent decides whether to relocate:

- It generates a random number **p**, uniformly distributed on [0, 1).

- If $u_a(h) < 1$ or ($u_a(h) = 1$ and **p < m**), then it tries to relocate, otherwise it stays at **h**. That is, an agent tries to relocate in two cases: (a) when it is dissatisfied or (b) at probability **m** if satisfied.

Step 2: If agent **a** decides to relocate, then it searches for a new location and decides whether to move:

- It constructs a set $V_a$ of opportunities for potential relocation by randomly selecting **w** unoccupied cells from all cells that are unoccupied at that moment.

- It estimates a utility $u_a(v)$ of each $v \in V_a$ and selects the one with the highest utility $u_a(v_{best})$. If there are several best vacancies in $V_a$, it chooses one of them randomly.

- It moves to $v_{best}$ if either of the following two conditions are met (otherwise it stays at **h**):
    - $u_a(h) < 1$ and $u_a(v_{best}) > u_a(h)$, i.e., it is dissatisfied with its current cell **h** and $v_{best}$ is better than **h**.
    - $u_a(h) = 1$ and $u_a(v_{best}) = 1$, i.e., it is satisfied with **h** but moves to $v_{best}$ which is also satisfactory.

At every time step, each agent performs the above steps once. Following Schelling's original approach, we use asynchronous updating (Cornforth et al. 2005) and allow each agent to make its decision based on the instantaneous state of the pattern. Agents are processed in random order, which is established anew at each time step.

Our model rules differ from Schelling's description in the following respects:
- A satisfied agent tries to relocate with nonzero probability **m.** Schelling's assumption is **m** = 0.
- The distance between cells has no effect on agents' relocation. In Schelling's model, agents move to the closest satisfactory position.
- Agents can move from one unsatisfactory cell to another if the relocation increases the number of neighboring friends. In Schelling's description, agents can only move to cells that they find satisfactory[2].

## 5. Model Investigation

*5.1. The tolerance thresholds of agents*

We explore different distributions of agents' tolerance thresholds in order find the ones that produce mixed patterns. We limit our investigation to symmetric cases where the tolerance distribution within each color group is identical.

We examine two families of discrete distributions:

---

[2] In one of the variations of the linear model, where agent movement is restricted by distance, Schelling (1971, p153-154) formulated a similar mechanism where agents would move to a neighborhood with 3/8 friends, if no nearby neighborhood with the desired 4/8 exists.

1. Dichotomous: Blue and green populations consist of two subgroups: The tolerance threshold is $F_1$ for a fraction **α** of agents (of each color) and $F_2$ for the rest.
2. Beta-binomial: It is sufficiently flexible for generating a variety of tolerance threshold distributions.

We explore different fractions of blue and green agents as well. We denote **β** as the fraction of blue agents, e.g. **β** = 0.5 indicates equal numbers of blue and green agents.

We use a 5×5 Moore neighborhood, which results in 25 values of **F** ∈ {0/24, 1/24, …, 24/24} for the majority of the situations[3]. We choose a 5x5 neighborhood because the classical 3×3 neighborhood masks the fine details of the model's dependency on **F**, as it entails only nine values of **F**.

*5.2 General settings*

Unless otherwise stated, we use the following settings in all of the experiments:
- Grid dimensions of 50×50 cells.
- 98% of the grid is occupied, that is, 50 cells are vacant.
- Equal number of blue and green agents (**B**=0.5)
- An agent evaluates **w** = 30 unoccupied cells per time step when considering relocation.
- The probability of a relocation attempt by a satisfied agent is **m**=0.01.
- The initial color and tolerance patterns are random.

## 6. Evaluation of the model outcomes

We estimate the long-term behavior of the model by running it for 50,000 time steps. For most of the experiments presented in this study, the model patterns converge to a quasi-stable state in which their statistical properties do not change. This quasi-stable state is independent of the initial spatial configuration of agents.

The model's patterns can stall when none of the agents have suitable relocation opportunities. For the settings used in this study (section 5.2) and a homogeneous tolerance threshold **F** for all agents, the patterns stall at a highly segregated configuration when **F**≥14/24.

---

[3] In practice, more than 25 values of the fraction of friends are possible because of empty cells. However, 25 discrete values are sufficient for the high density of agents that we study.

The model produces two patterns: a pattern of color (i.e., blue and green agents) and a pattern of the agents' tolerance thresholds. We follow Hatna and Benenson (2012) in characterizing these patterns using the three measures listed below.

*6.1. Measuring segregation by color ($I_C$) and tolerance ($I_T$)*

The level of segregation of the color and tolerance pattern is measured using the Moran's **I** index of spatial association (Anselin 1995). We denote $I_C$ as the index for color and $I_T$ as the index for tolerance segregation. For random (integrated) patterns, **I** is close to zero (Figure 1a), while for fully segregated patterns, **I** is close to 1 (Figure 1b)[4].

An intermediate value of $I_C$ implies a moderate level of color segregation, but it does not necessarily indicate the presence of a mixed pattern. In order overcome this limitation of $I_C$, we introduce the **C** index.

*6.2. Identifying mixed patterns (**C** index)*

We employ the **C** index to identify mixed patterns. The index evaluates how closely a color pattern resembles a mixed pattern with three equal sized (green, blue and integrated) patches, such as the one shown in Figure 1c.

The index is based on partitioning the color pattern into three regions: (1) pure blue, (2) pure green and (3) integrated. The partition includes the boundary between the three regions as well (Figure 3). We define the **C** index as the size of the smallest region relative to the size of the entire grid. A region that is not present in the pattern has a size of zero.

Figure 3 depicts the partition of the mixed pattern shown in Figure 1c into the three regions. The three regions are of similar size, and each takes up about 21% of the total area, hence **C**=0.21. C is zero for a pattern that do not contain all of the three regions, such as the integrated and segregated patterns shown in Figures 1a and 1b. A detailed description of the **C**-index algorithm can be found in Hatna and Benenson (2012).

---

[4] For the specific parameters used in this study (density of **d**=0.98 and neighborhoods dimensions of 5x5), the highest value of $I_C$ is about 0.93.

The three indices are calculated based on 5×5 Moore neighborhoods. All indices presented below are averaged over 30 runs of the model.

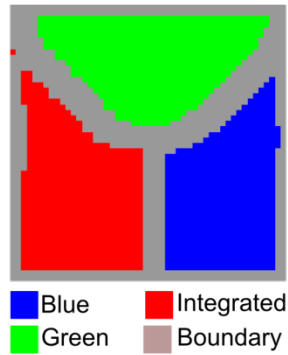

Figure 3: The partition of the mixed pattern shown in Figure 1c into blue, green and integrated patches and a boundary.

## 7. Results

We first review the classic case of common **F** and demonstrate its dependence on **m**. Then, we explore the patterns produced by dichotomous tolerance thresholds. Finally, we investigate the patterns for a beta-binomial distribution of the agents' tolerance.

*7.1. Cases of common tolerance threshold for all agents*

We consider the case where all agents have the same tolerance threshold **F** for demonstrating the dependence of the model's steady patterns on the rate **m** of relocation attempts by satisfied agents.

7.1.1. The case where satisfied agents do not move (**m**=0)

For **m** = 0, the steady pattern is very sensitive to the initial one. If the initial pattern is fully segregated, then it remains unchanged for all **F**. If the initial pattern is random (i.e., fully integrated), then it remains random for **F** ≤ 7/24 but converges to segregation for **F** ≥ 9/24. For the intermediate case of **F** = 8/24, an initially random patterns stalls, after some 10 time steps in a partly segregated state, which is essentially dependent on the initial pattern.

7.1.2 The tipping point depends on **m** (**m**>0)

When **m**>0, the tipping point between integration and segregation depends on **m** (Figure 4). For **m** ∈ [0.01,0.04], the transition between integration and segregation occurs at $F_{critical}$ = 5/24. For **m** ∈ [0.05, 0.69], it occurs at $F_{critical}$ = 6/24, and for **m** ∈ [0.70,1.00] at $F_{critical}$ = 7/24. This increase of $F_{critical}$ is related to the disturbing influence of random migration during the initial clustering of agents.

In what follows, we investigate the Schelling model for **m** = 0.01 wherein $F_{critical}$ = 5/24.

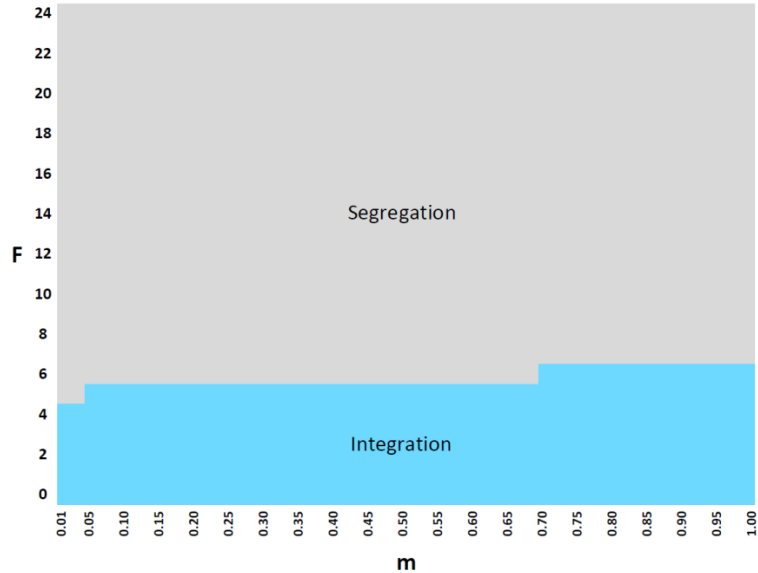

Figure 4: The domains of integration (blue) and segregation (grey) as dependent on the tolerance threshold **F** and the rate **m** of relocation attempts by satisfied agents.

7.1.3 The time of convergence to segregation depends on **F**

Given **m** = 0.01, an initial random pattern converges to segregation for **F** ≥ 5/24. The time of convergence depends on **F** (Figure 5). For $F_{critical}$=5/24, it takes about 10,000 time steps, but for higher tolerance values such as **F**≥12/24, it takes no more than 15 time steps.

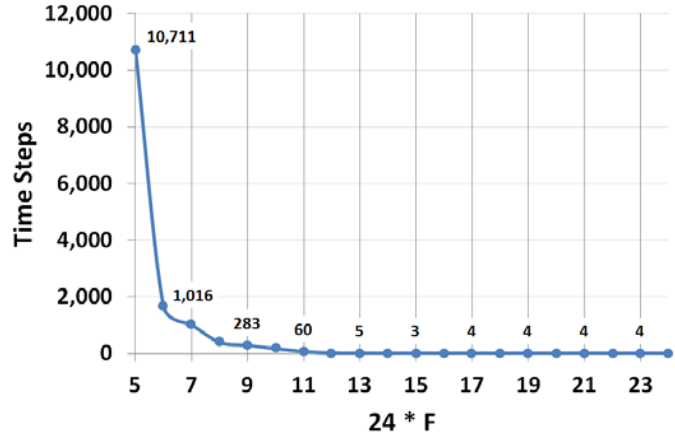

Figure 5: The average number of time steps required to reach $I_c = 0.8$, starting with a random pattern as dependent on **F** (for **m** = 0.01, averaged over 10 runs).

Figure 6 shows the dynamics of the $I_c$ and **C** indices and the steady model patterns for **F**=5/24 and **F**=8/24. Patches of green and blue agents emerge and grow until two large patches, one for each group, are formed. Afterwards, the length of the boundary between the two patches converges to a minimum. Figure 6 illustrates that the process is slower for the critical value of **F**=5/24 as compared with **F**=8/24. It is worth noting that during the growth of the blue and green patches, the patterns contain integrated areas as well. Such patterns are mixed according to the value of **C**, but they dissolve as the blue and green patches take over the entire pattern.

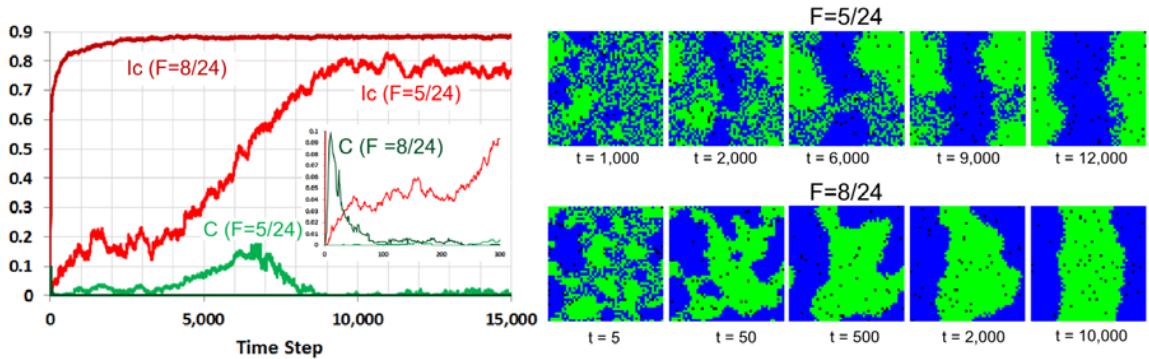

(a) Dynamics of $I_c$ and **C**  (b) Color patterns at different time steps

Figure 6: The dynamics of two model runs: **F** = 5/24 and **F** = 8/24.

*7.2. Two equal sized groups of different tolerance thresholds*

Up to this point, all agents had the same tolerance thresholds. We now pay attention to cases where each color group is composed of two equal sized subgroups: half of blue and green agents ($\alpha$=0.5) have a tolerance threshold $F_1$, while the other half has a tolerance threshold $F_2$. We keep the number of blue and green agents equal as well ($\beta$=0.5).

We study the model dynamics for four values of $F_1$: (a) zero – 0/24, (b) below the tipping point – 3/24, (c) tipping point – 5/24, (d) above the tipping point – 7/24 and then present the general behavior for all ($F_1$, $F_2$) pairs. We conclude this section by demonstrating the temporal development of the mixed patterns.

To remind, we average all numeric results for a given set of model parameters over 30 model runs. The steady patterns are considered at $t$ = 50,000.

### 7.2.1. $F_1$ = 0, $F_2$ varies

We begin with the case where half of the agents of each color are fully tolerant $F_1$ = 0/24. These agents are satisfied within any neighborhood and attempt to relocate only for random reasons at a rate $m$. We investigate the model patterns as dependent on the tolerance $F_2$ of the other half of the agent population.

Figure 7a presents the dependence of the three measures (Moran's $I_C$ for color, Moran's $I_T$ for tolerance and $C$ for detecting mixed color patterns) on $F_2$. When $F_2$<5/24, the pattern is integrated. Starting at $F_2$ = 5/24, the value of $I_C$ increases linearly until it reaches a value of 0.46 and stabilizes at $F_2$ = 18/24. $I_T$ is non-zero from $F_2$ = 12/24, and then increases up to 0.7, exceeding $I_C$ at $F_2$ = 15/24. $I_T$, just as $I_C$, stabilizes at $F_2$ = 18/24. $C$ clearly indicates that for $F_2 \geq$ 18/24, the color pattern is mixed (e.g. Figure 7b, $F_2$ =20/24).

Mixed patterns appear when the tolerance patterns become segregated. Blue and green patches consist of intolerant $F_2$ agents, while the integrated area is made of the completely tolerant $F_1$ agents.

When $F_2 \leq$12/24, the pattern of tolerance is integrated because $F_2$ agents are content to reside within integrated areas or within the boundary of segregated patches of their own color, where about half of their neighbors are friends, while the completely tolerant $F_1$ agents are satisfied within any neighborhood. However, for $F_2$>12/24, segregation by tolerance is observed as $F_2$ agents seek local majority and therefore tend to leave diverse areas, while the $F_1$ agents are willing to stay within integrated areas.

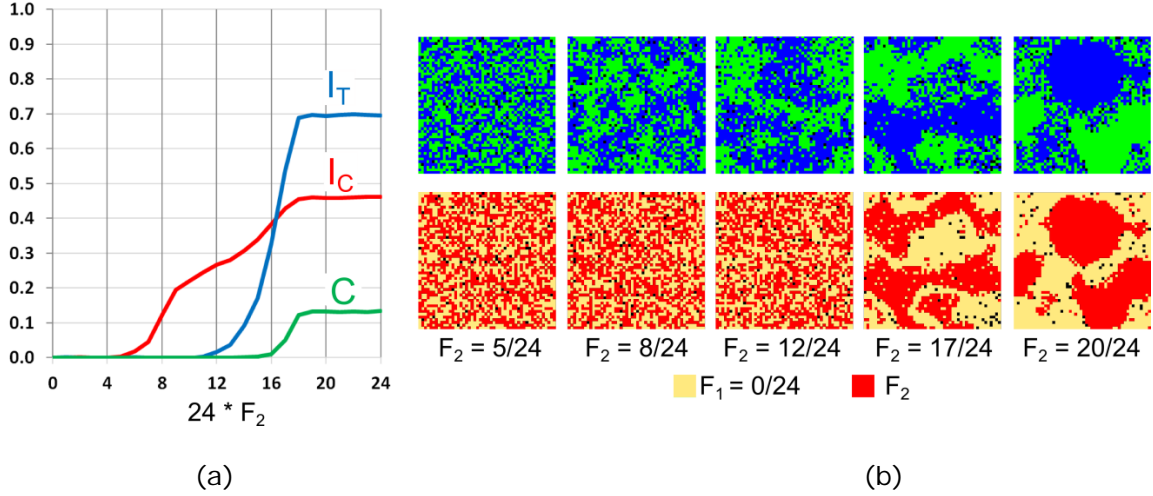

(a)  (b)

Figure 7: The case of $F_1$=0/24 and varying $F_2$: (a) Moran's I for agents' color ($I_c$), tolerance ($I_T$), and the C-index. (b) Color and tolerance patterns at t = 50,000.

7.2.2. $F_1$ = 3/24, $F_2$ varies

When $F_1$ = 3/24, the $F_1$ agents slightly react to neighbors. The color pattern is mixed for high values of $F_2$, however the dependence of the patterns on $F_2$ is different form the case of $F_1$ = 0 (Figure 8).

For $F_2$≤6/24 (and not 5/24, as above), the color pattern is integrated. Starting from $F_2$=7/24, $I_c$ gradually increases as the integrated area is replaced by the growing blue and green patches until it reaches a maximum of 0.7 at $F_2$=12/24. At this value of $F_2$, the steady pattern consists of two segregated patches and only the boundary between the patches is integrated (Figure 8b, $F_2$ = 12/24).

The value of $I_T$ begins to increase at $F_2$=9/24, since $F_1$ agents concentrate at the boundaries of the green and blue patches (Figure 8b, $F_2$=12/24). $I_c$ begins to drop at $F_2$=13/24 while $I_T$ continues to increase, indicating that the segregated pattern is replaced by a mixed one. The three indices stabilize at $F_2$=16/24, and C indicates that the patterns are mixed (C ~ 0.15).

Note that within the interval 8/24 ≤ $F_2$ ≤ 15/24, the model patterns are mixed, albeit the area of integration is smaller (0.05 < C < 0.1), and the level of segregation of the tolerance patterns is very low. We will show that this type of pattern is characteristic for $F_1$ = 3/24 in Section 7.2.5.

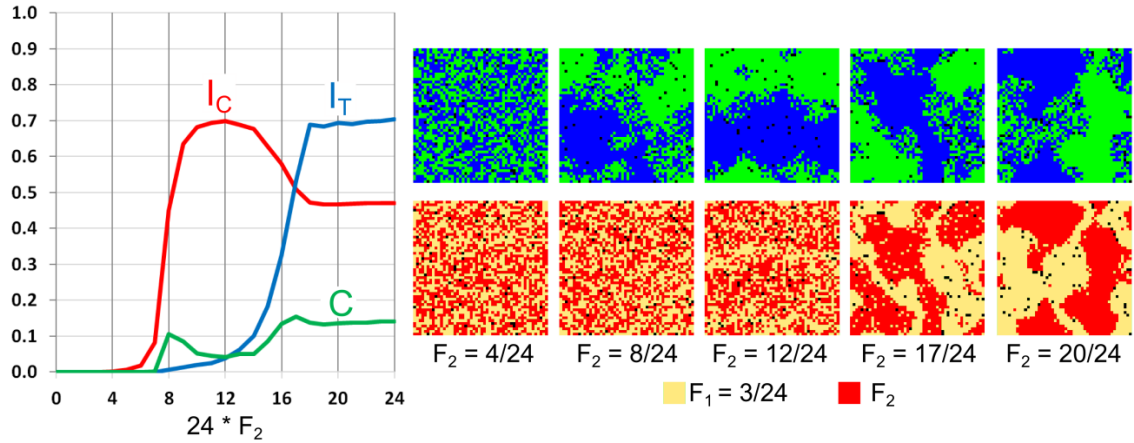

Figure 8: The case of $F_1$ = 3/24 for varying $F_2$: (a) Moran's I for agents' color ($I_c$) and tolerance ($I_T$), and the **C**-index. (b) The patterns of agents' color and tolerance at **t** = 50,000.

7.2.3. $F_1$ = 5/24, $F_2$ varies

For $F_1$ = 5/24 (i.e., $F_1$ equals to the tipping point), an abrupt transition from integration to segregation occurs, as expected, at $F_2$ = 5/24 (Figure 9). Segregation by tolerance starts at $F_2$=13/24, indicating that tolerant agents begin to concentrate at the boundaries between the segregated patches (Figure 9b), but the patterns remain segregated until $F_2$ =17/24. At $F_2$ =18/24, the integrated patch appears and it reduces the level of color segregation ($I_c$), while the level of tolerance segregation grows rapidly. By $F_2$ =19/24, the mixed patterns are fully formed and the indices stabilize.

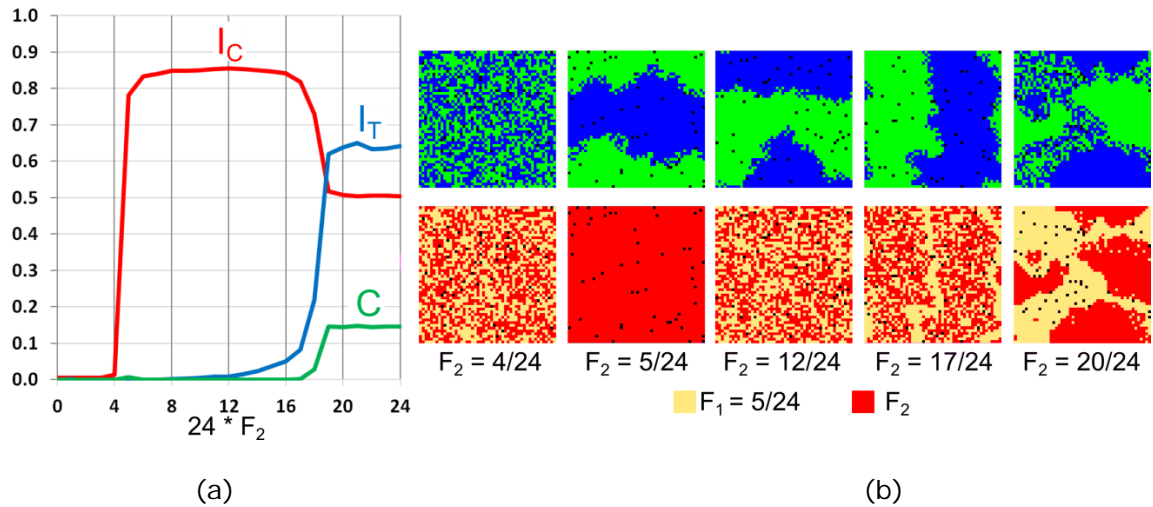

(a)                                                                (b)

Figure 9: The case of $F_1 = 5/24$ and varying $F_2$: (a) Moran's I for agents' color ($I_c$) and tolerance ($I_T$), and **C**. (b) The patterns of agents' color and tolerance at $t = 50,000$.

7.2.4. $F_1 = 7/24$, $F_2$ varies

For $F_1 = 7/24$ which is above the tipping point, the model exhibits a simple integration – segregation dichotomy and mixed patterns do not emerge. Segregation is reached abruptly at $F_2 = 3/24$ and segregation by tolerance begins at $F_2 = 13/24$, with the $F_1$-agents located along the border of the blue and green patches.

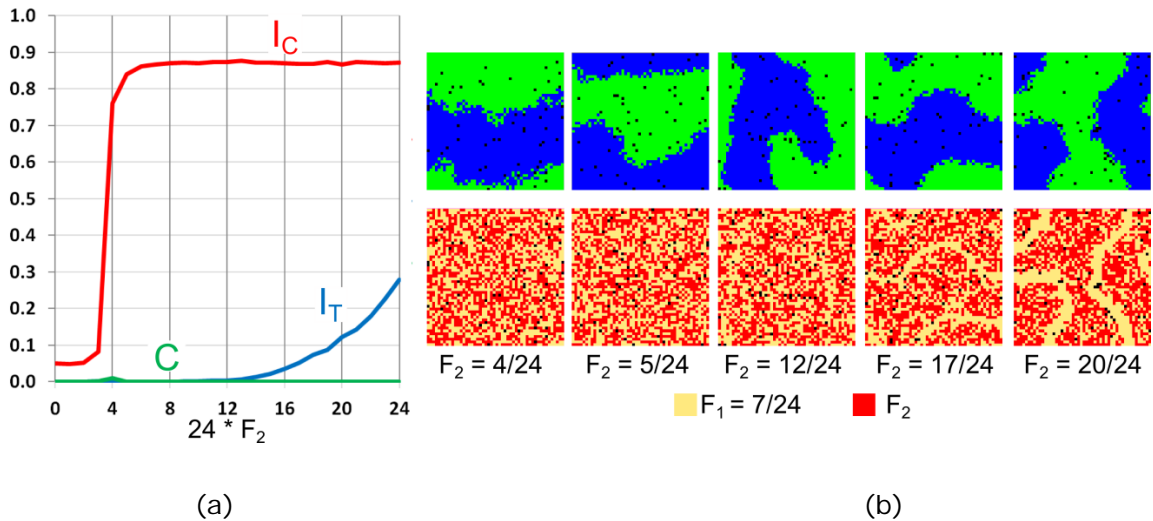

(a)                                                                (b)

Figure 10: The case of $F_1 = 7/24$ and varying $F_2$: (a) Moran's **I** for agents' color ($I_c$) and tolerance ($I_T$), and **C**. (b) The patterns of agents' color and tolerance at $t = 50,000$.

## 7.2.5. Varying $F_1$ and $F_2$

The case where both $F_1$ and $F_2$ vary ($\alpha = 0.5$, $\beta = 0.5$) is presented in Figure 11. The heat map of **C** (Figure 11c) confirms that a mixed pattern emerges when the tolerance of one of the subgroups is at a tipping point of 5/24 or below, while the tolerance of the other subgroup is 16/24 or above. We also see that in these cases, the tolerance patterns are segregated (Figure 11b).

For $F_1 \in \{0, 1/24, 2/24\}$, segregation by color ($I_C$) increases gradually with the increase of $F_2$ until a mixed pattern is formed. This is indicated by the emergence of segregation by tolerance ($I_T$) at $F_2 = 13/24$. Segregation by tolerance exceeds segregation by color at $F_2 = 17/24$.

For $F_1 \in \{3/24, 4/24, 5/24\}$, with the increase of $F_2$, the pattern become segregated before a mixed pattern emerges. Mixed patterns do not emerge for cases of $F_1 > 5/24$ that are below the diagonal.

The heat maps show cases where mixed patterns are formed without segregation of tolerance. This behavior is limited to cases, such as $F_1=3/24$, $F_2 \in \{8/24,..,14/24\}$ and at $F_1=4/24$, $F_2=6/24$ (Figure 11c).

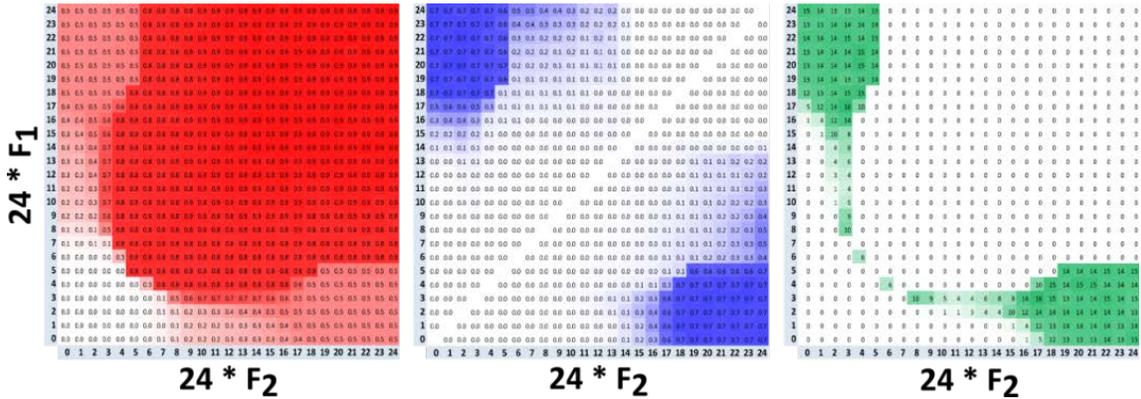

(a) Moran's **I** for color  (b) Moran's **I** for tolerance  (c) **C** in percentages

Figure 11: Heat maps for the three indices for all pairs of $F_1$ and $F_2$ at $t=50,000$.

### 7.2.6. The formation of mixed patterns

We now demonstrate how the mixed patterns emerge, using the case of $F_1=3/24$ and $F_2=20/24$. Figure 12a depicts the dynamics of a model run that begins with a fully segregated pattern of color.

The level of color segregation ($I_C$) of the model pattern decreases, while the level of segregation by tolerance ($I_T$) and the **C**-index increase. Visually, the boundary between the segregated parts of the pattern dissolves towards time step 100 and then, between time steps 200 and 500, tolerant $F_1$ agents form a steady integrated area.

The fraction of migrating $F_1$ agents, who are highly tolerant, remains close to **m** during the entire run (Figure 12b). The relocation rate of intolerant $F_2$ agents is higher than the one of the $F_1$-agents at the first 200 time steps when $F_2$ agents near the boundary of the segregated parts of the pattern relocate into the internal parts of the patches. These migrations reduce the number of empty cells within the segregated patches, while increasing their number within the integrated boundary area (see patterns for $t > 200$). As a result, cells that are suitable for the $F_2$ agents become rare, and the actual migration rate of the $F_2$ agents decreases far below **m**, starting from $t = 200$ (Figure 12a, right).

Figure 12b represents the model dynamics for the same tolerance distribution ($F_1=3/24$ and $F_2=20/24$) using a random initial pattern. At $t=0$, all of the $F_1$ agents are satisfied, while all of the $F_2$ agents are not. As a result, more than 60% of the $F_2$ agents relocate during the first time step, and during the next 50 time steps, the value of the $I_C$ grows. The migration rate of the $F_1$-agents remains very close to the value of **m** (1%). As in the case of a segregated initial pattern, $F_2$-agents tend to avoid the boundaries and relocate to the internal parts of the patches of their own color. This initiates the segregation by tolerance that is reflected by the growth of $I_T$ and **C**. The global characteristics of the patterns for the initially segregated and random conditions become identical towards $t = 200$, while the model patterns remain visually different for a longer period.

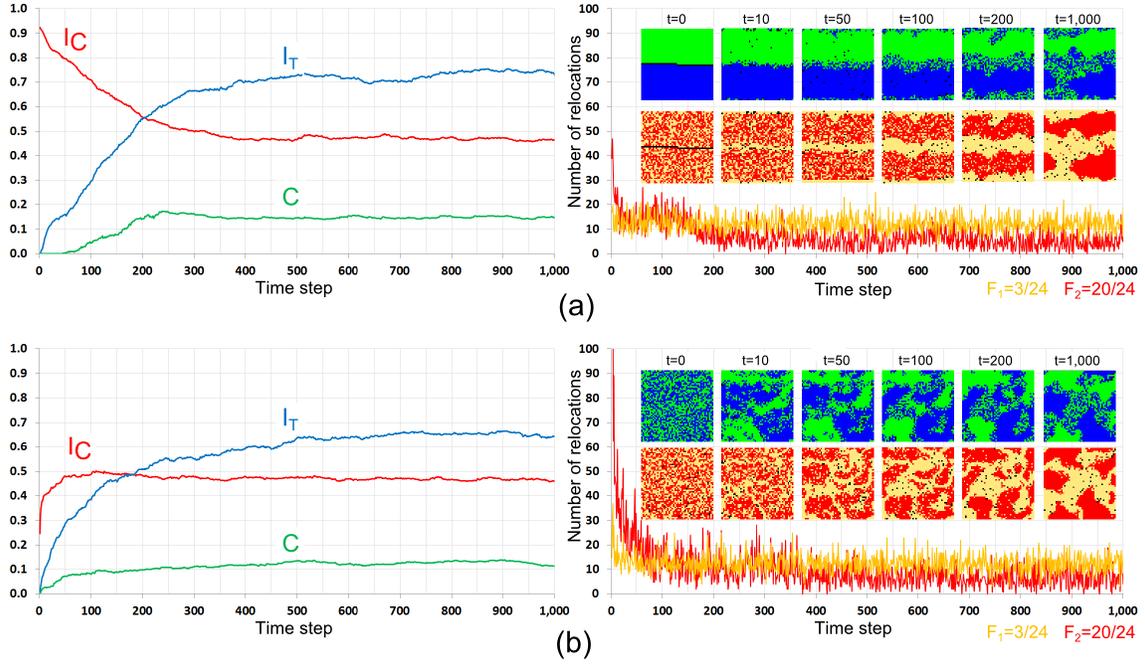

Figure 12: The formation of a mixed pattern for the case of **F₁**=3/24 and **F₂**=20/24, starting from a (a) random and a (b) segregated initial pattern.

### 7.3. Varying *α*, *β* and *m*

In the previous section, the number of **F₁** and **F₂** agents (**α**=0.5) was equal, as was the number of blue and green agents (**β**=0.5), while the rate of relocation attempts by a satisfied agents was low (**m**=0.01). In this section, we explore how **α, β** and **m** influence the mixed patterns.

7.3.1. Varying fraction **α** of **F₁** agents

We consider the same case of **F₁** = 3/24, **F₂** = 20/24, but for varying fraction **α** of **F₁**-agents (Figure 13).

The size of the integrated patch is almost proportional to **α.** When **α** is close to zero, the pattern is segregated, because highly intolerant **F₂** agents comprise the vast majority of the population. In this case, the **F₁** agents are located at the boundary separating the blue and green patches (Figure 13b, **α**=0.05**)**. As **α** increases**,** the size of the homogenous patches decreases, while the integrated area grows, as reflected in the steady decrease of **I_C**. The level of segregation by tolerance (**I_T**) and the size of the smallest patch **C** both grow, with the increase of **α** up to **α** = 0.5. At **α** = 0.5, the size of the three regions is similar and $I_T$ and **C** are highest. With further increase of **α**, the homogenous patches continue to shrink and the integrated area dominates.

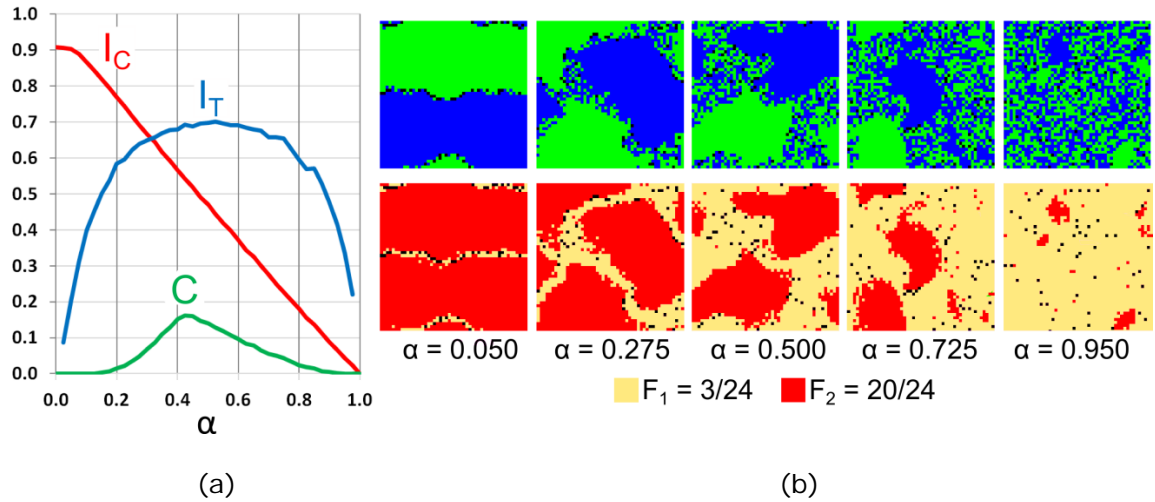

(a)                        (b)

Figure 13: Varying fraction **α** of $F_1$ agents for $F_1$=3/24, $F_2$=20/24: (a) Moran's I for agents' color ($I_c$), tolerance ($I_T$) and **C**. (b) The patterns of agents' color and tolerance at **t** = 50,000.

7.3.2. Varying fraction **β** of blue agents

We now explore how the fraction **β** of blue agents affects the mixed patterns for the same case of $F_1$=3/24, $F_2$=20/24, but we keep the fraction of $F_1$ and $F_2$ agents equal (**α**=0.5) (Figure 14).

When the vast majority of agents are green (Figure 14b, **β** = 0.05), the few blue agents form a small patch. The blue $F_2$ agents reside within the patch, while the blue $F_1$ agents occupy the edge but are too few to create a distinct integrated area. As **β** increases, the integrated area becomes apparent and grows until reaching a maximum at **β** = 0.5. $I_T$ and **C** increase as **β** grows to 0.5 and then decrease, while $I_c$ behaves in the opposite manner.

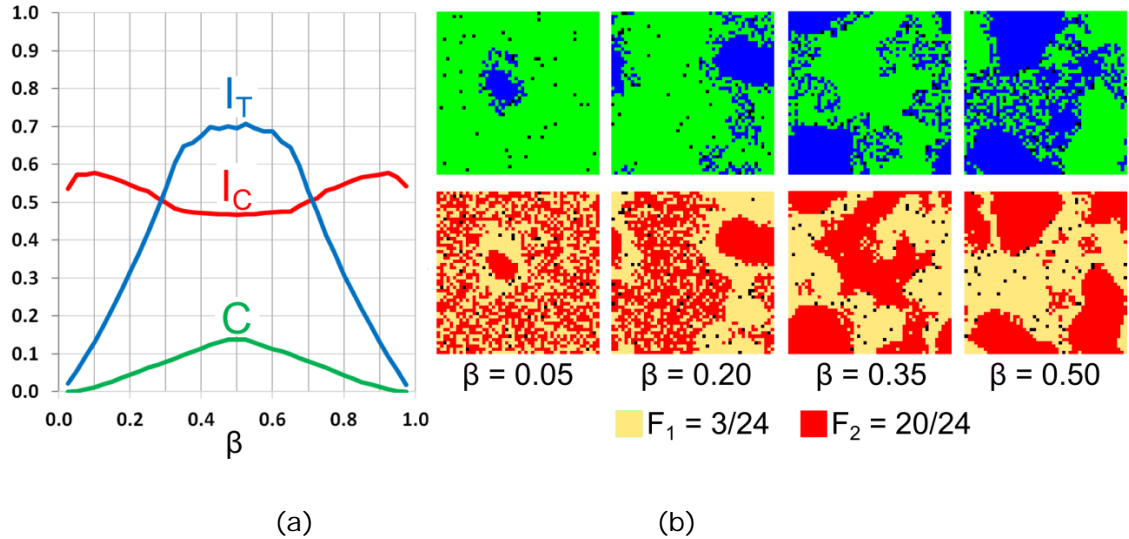

(a) (b)

Figure 14: The case of varying fraction **β** of blue agents for **$F_1$**=3/24, **$F_2$**=20/24: (a) Moran's I for agents' color (**$I_c$**), tolerance (**$I_T$**). (b) The patterns of agents' color and tolerance at **t** = 50,000.

7.3.3. The dependence of the mixed patterns on **m**

To investigate the dependence of the mixed patterns on the probability **m** that a satisfied agent would attempt to relocate, we consider the case of **$F_1$** = 3/34 and vary **$F_2$** and **m** (Figure 15).

According to the heat maps, given **$F_2$** ≥ 18/24, mixed patterns emerge for any **m**≥0.01. For these values of **$F_2$**, the tolerance pattern is always segregated (Figure 15b) and the value of **C** is high (Figure 15c).

The heat maps also indicate that for a **$F_2$** between approximately 8/24 and 17/24, the model produces a mixed pattern without the segregation of agents by tolerance. These patterns are sensitive to the value of **m** and tend to diminish as **m** grows.

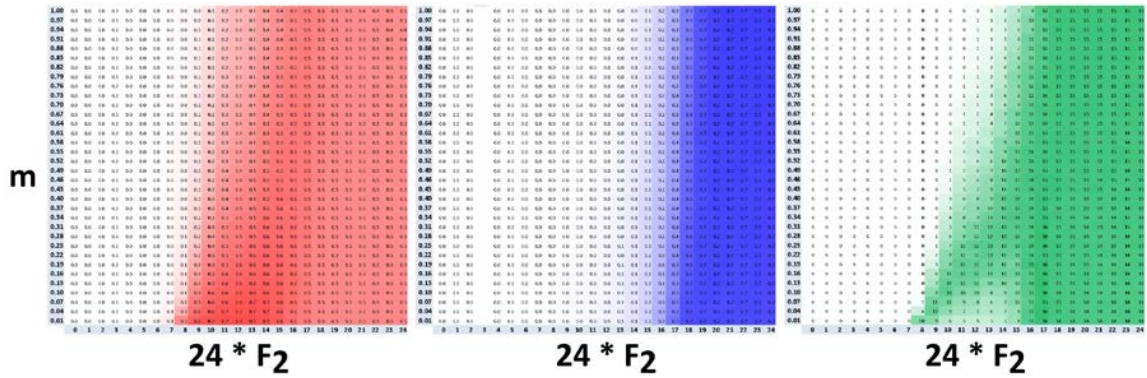

(a) Moran's **I** for color    (b) Moran's **I** for tolerance    (c) **C** in percentages

Figure 15: The dependence of the three indices on **m** and $F_2$ for the case of $F_1$=3/24 and **α, β**=0.5.

7.3.4. The dependence of the mixed patterns on neighborhood size

We now test how neighborhood size affects the mixed patterns. We limit the investigation to the case of $F_1$ = 0/24, $F_2$ = 20/24 and vary the neighborhood size from 3x3 to 9x9. In order to keep the neighborhood small relative to the grid size, we use a twice as large grid of 100x100 cells. We also increase the number of empty cells **w** considered by agents when relocating from 30 to 120.

For 3x3 neighborhoods, the intolerant $F_2$ agents produce many small homogeneous patches while the tolerant $F_1$ agents occupy the boundaries and produce small integrated areas (Figure 16). For larger neighborhood sizes, the agents produce mixed patterns that are more distinct. These patterns consist of two large homogeneous patches and an integrated area.

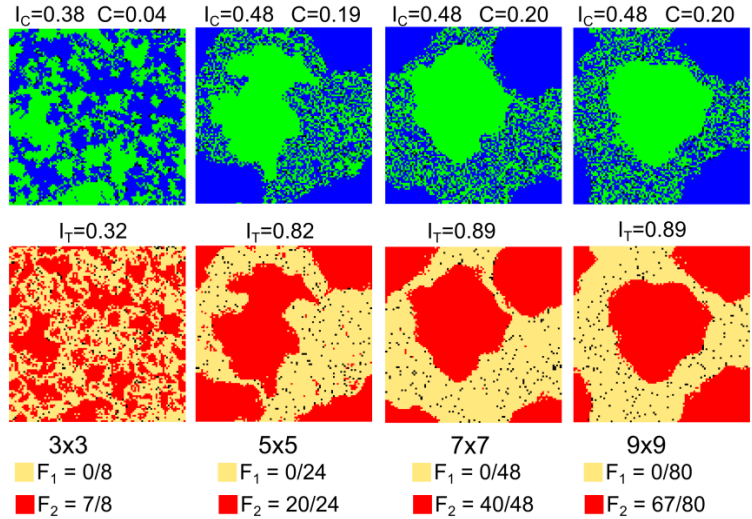

Figure 16: Steady mixed patterns for different neighborhood sizes. All indices are calculated based on 5x5 neighborhoods.

*7.4. Beta-binomial distribution of tolerance*

To go beyond dichotomous distributions of tolerance, we consider cases where agents of both colors are assigned one of 25 ({0/24, 1/24, …, 24/24}) tolerance thresholds according to a beta-binomial distribution. We keep the number of blue and green agents equal (**β**=0.5). We compare five qualitatively different cases by varying the distribution parameters (Figure 18).

The first case is a positively skewed unimodal distribution wherein the tolerance of about 1/3 of the agents is below the tipping point (**F** < 5/24) and the average is about 7/24 (Figure 18a). This distribution leads to a pattern that is mixed by color and integrated by tolerance (**C**=0.07, $I_C$ = 0.49, $I_T$ = 0.02).

The second case is a symmetric unimodal distribution (Figure 18b) where only about 4% of the agents have a tolerance threshold below the tipping point and the average is 12/24, which is much higher than the tipping point. As expected, the steady color pattern is segregated, while the tolerance pattern remains integrated, $I_C$ =0.84, $I_T$ =0.03, **C** = 0.

The third case is a uniform distribution (Figure 18c) with 20% of the agents having tolerance thresholds below the tipping point and an average of 12/24. The steady pattern in this case is segregated by color ($I_c$ = 0.70), with the small integrated patches within the larger segregated areas (**C** = 0.07). The tolerance pattern is somewhat segregated (**I$_T$** = 0.16) with a clear concentration of the tolerant agents on the boundary between the color patches.

The fourth case is a symmetric U-shaped distribution with about 29% of the agents having tolerance thresholds below the tipping point (Figure 18e), with an average of 12/24 (Figure 18e). The color pattern is somewhat mixed (**I$_C$** = 0.52, **C** = 0.09), while the segregation of tolerance is high (**I$_T$** = 0.45) but still below **I$_C$**.

The fifth case is another symmetric U-shaped distribution with higher variance. The average tolerance is 12/24 and 39% of the agents have tolerance thresholds below the tipping point. This leads to a mixed color pattern that looks more distinct (**C** = 0.10) and a level of segregation by tolerance (**I$_T$**=0.61) that is higher than the level segregation by color (**I$_C$**=0.46).

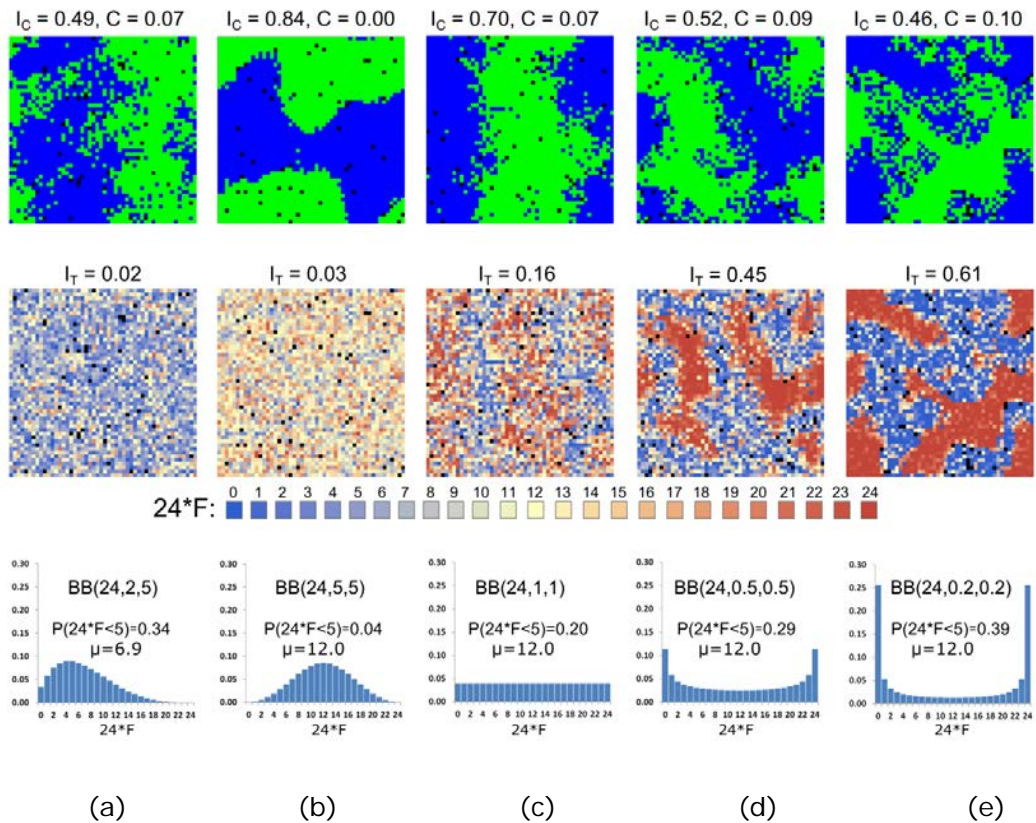

(a)  (b)  (c)  (d)  (e)

Figure 18: Steady color (top) and tolerance (middle) patterns for five distributions of tolerance thresholds (bottom) at **t** = 50,000.

From this limited investigation, we can hypothesize that mixed patterns become more distinct as the variance of the tolerance distribution increases and the vast majority of agents are either highly tolerant or highly intolerant.

## 8. Summary

Stemming from ethnic urban patterns, which in many cases are composed of homogeneous and heterogeneous neighborhoods, we investigated ethnic residential dynamics for a population of individuals that differ in their tolerance to other ethnic groups. Formally, we studied the Schelling model dynamics for a heterogeneous population of agents.

Most of our investigation focused on dichotomous cases, where the level of agents' tolerance is either $F_1$ or $F_2$. We found that in these cases, the model produces mixed patterns when the tolerance of the $F_1$ agents is equal to or below the tipping point ($F_1 \leq 5/24$), while agents of the second group need more than 2/3 of their neighbors to be friends ($F_2 \geq 16/24$). For these tolerance distributions, the $F_1$-agents are satisfied almost everywhere, while the $F_2$-agents are highly intolerant and seek neighborhoods with sufficiently high fractions of friends. The steady pattern of agents' tolerance in these cases is segregated and the blue and green $F_2$-agents form two segregated patches, while the $F_1$ agents of both colors concentrate within the integrated areas that separate the patches of the $F_2$-agents.

When we set $F_1 \leq 1/24$ and investigated the model patterns, varying $F_2$ from 0/24 to 24/24, we found that an increase of $F_2$ causes a gradual transformation from integration to mixed. However, for $F_1 \in [3/24, 5/24]$, with the growth of $F_2$, the pattern first changes from integration to segregation and only then from segregation to mixed.

Mixed patterns emerge because intolerant agents leave integrated areas and create dense blue and green patches with few empty cells. Tolerant agents stay within the integrated areas and migrate within them for random reasons, but are not able to enter the dense segregated patches. Integrated areas thus serve as extended boundaries between segregated patches.

As long as a part of the population is tolerant, while the rest is highly intolerant, mixed patterns emerge regardless of other model parameters. In particular, the patterns remain mixed irrespective of the fraction $α$ of $F_1$-agents, fraction $β$ of blue agents, the rate $m$ of random relocation attempts and the neighborhood size. Model parameters influence the size of the segregated and integrated patches, but the mixed nature of the patterns is always preserved.

If the agents' tolerance is non-dichotomous, but distributed over the entire [0, 1] range, mixed patterns became more distinct as the fraction of agents with extreme tolerance threshold increases.

The Schelling model produces mixed patterns due to the relocations of intolerant agents out of the integrated areas. Similar residential movements, such as the relocation of white residents out of ethnically integrated neighborhoods (Crowder 2000), might have contributed to the formation of mixed patterns in cities. However, models that include additional factors, such as economic status and dwelling prices (Benard and Willer 2007; Fossett 2006; Hatna and Benenson 2011), are likely to provide additional insight into the formation of mixed patterns as segregated and integrated neighborhoods may attract different segments of the residents' population of the same ethnicity.

## 9. Acknowledgements

Erez Hatna acknowledges support from J. M. Epstein's NIH Director's Pioneer Award, number DP1OD003874 from the National Institute of Health.